\documentclass[10pt,technote]{IEEEtran}
\usepackage{graphicx}
\usepackage[ruled]{algorithm2e}
 
\usepackage{indentfirst}
\usepackage{graphicx}              
\usepackage{subcaption}            
\usepackage{graphicx}
\usepackage{bm}
\usepackage{amsmath}
\allowdisplaybreaks[4]
\usepackage{amssymb}
\usepackage{times}
\usepackage{mathtools}
\usepackage{psfrag}
\usepackage{cite}
\usepackage{lastpage}
\usepackage{fancyhdr}
\usepackage{color}
\usepackage{amsthm}
\usepackage{bigints}
\usepackage{multirow}
\usepackage{booktabs}
\usepackage{threeparttable}
\usepackage{array}
\usepackage{threeparttable}
\usepackage{xcolor}
\usepackage{soul}
\usepackage{amsmath}
\usepackage{mathtools}
\usepackage{makecell}
\theoremstyle{remark}

\begin{document}
\title{On the Optimal Power Allocation and Performance Analysis for the HSIC-PA Aided Hybrid NOMA}

\author{
Ning Wang, Ziyue Lv, Yanshi Sun,~\IEEEmembership{\itshape Member, IEEE}, Minghui Min, ~\IEEEmembership{\itshape Senior Member, IEEE}
and Shiyin Li
\thanks{
The work of Ning Wang was supported in part by the National Key Research and Development Program of China under Grant 2024YFC3015801, in part by Xinjiang Talent Development Funds under Grant XJRC-2025-GX-YJ-QY-009, and in part by the Emergency Communications Equipment Innovation ``Unveiling the List and Taking the Lead" Initiative under Project E04-29. The work of Ziyue Lv was supported in part by Postgraduate Research and Practice Innovation Program of Jiangsu Province under Grant 26CXJH4819, and in part by Graduate Innovation Program of China University of Mining and Technology under Grant 2026WLJCRCZL359. The work of Yanshi Sun was supported in part by the National Natural Science Foundation of China under Grant 62301208 and in part by the Hefei University of Technology Academic Novice Advancement Program B under Grant JZ2025HGTB0224.

Copyright (c) 2026 IEEE. Personal use of this material is permitted. However, permission to use this material for any other purposes must be obtained from the IEEE by sending a request to pubs-permissions@ieee.org. 

N. Wang, Z. Lv, M. Min and S. Li are with the School of Information and Control Engineering, China University of Mining and Technology, Xuzhou, 221116, China. (email: wangnsky@cumt.edu.cn, lvzy@cumt.edu.cn, minmh@cumt.edu.cn, lishiyin@cumt.edu.cn). N. Wang is also with Leon Technology Co.,Ltd, Urumqi, 830000, China.

Y. Sun is with the School of Computer Science and Information Engineering, Hefei University of Technology, Hefei, 230009, China. (email: sys@hfut.edu.cn). Corresponding author: Yanshi Sun.
}
}
\maketitle
\begin{abstract}
Hybrid non-orthogonal multiple access (H-NOMA), which leverages the complementary advantages of orthogonal multiple access (OMA) and NOMA, has emerged as a promising technique for next-generation wireless networks. This paper investigates power allocation optimization for the OMA and NOMA time slots allocated to the opportunistic user in H-NOMA systems, aiming to overcome the limitations of conventional equal power allocation strategies. Specifically, we consider an H-NOMA scheme that integrates hybrid successive interference cancellation (HSIC) with power adaptation (PA), referred to as the HSIC-PA aided H-NOMA scheme. To better adapt to practical communication scenarios, distinct power coefficients are assigned to the OMA and NOMA time slots of the opportunistic user to maximize the achievable data rate. Closed-form expressions for the optimal power allocation coefficients are derived, and simulation results validate the accuracy and effectiveness of the theoretical analysis.
\end{abstract}

\vspace{-1em}
\vspace{-1em}
\section{Introduction}

As wireless communications evolve towards the sixth generation (6G), next-generation multiple access (NGMA) technology has emerged as a research hotspot \cite{Liu2024_6G_Book}. Various advanced multiple access techniques, including non-orthogonal multiple access (NOMA) \cite{Liu2022_6G} and rate-splitting multiple access (RSMA) \cite{LinRSMA2021, LinRSMA2025}, have been investigated to improve spectrum utilization and system capacity. Among these techniques, NOMA has attracted considerable attention due to its high spectral efficiency and capability to support massive access. RSMA has also recently emerged as a generalized framework that bridges NOMA and space-division multiple access (SDMA) by splitting user messages into common and private streams. While RSMA offers flexibility in interference management, NOMA remains a low-complexity and practical solution, particularly when users experience distinct channel conditions and SIC can be reliably performed. In fact, NOMA can be regarded as a special case of RSMA when the entire message is treated as a private stream, which further motivates a thorough understanding of NOMA performance as a fundamental building block. Moreover, given the high compatibility and flexibility of NOMA with other 6G enabling technologies, such as reconfigurable intelligent surface (RIS) networks \cite{Kang2024_RIS, Lei2025_IRS}, integrated sensing and communications (ISAC) \cite{ISAC_2025}, and semantic communication (SemCom) \cite{SemCom_2025}, its potential for practical deployment is further enhanced.

Since most existing mobile networks are deployed with orthogonal multiple access (OMA) architecture, directly replacing OMA with NOMA is impractical. To address this issue, the recently proposed hybrid NOMA (H-NOMA) scheme \cite{Ding2019MEC} is a more feasible solution, which organically integrates NOMA and OMA. Users in H-NOMA are divided into legacy users and opportunistic users. Legacy users transmit only within their exclusively assigned time slots, while opportunistic users can transmit not only in their own allocated time slots (i.e., OMA transmission) but also in the time slots of legacy users by exploiting NOMA (i.e., NOMA transmission). The fundamental distinction between legacy and opportunistic users lies in their different QoS requirements. Legacy users prioritize transmission reliability, whereas opportunistic users aim to maximize data throughput. For example, a legacy user may represent an industrial monitoring sensor that periodically reports equipment status and requires reliable delivery of small data packets, whereas an opportunistic user may represent a mobile inspection robot that uploads image or diagnostic data to maximize its data throughput. 

In the process of NOMA transmission, multiple users share the same channel resource block for parallel transmission, which inevitably causes inter-user interference. To address this issue, successive interference cancellation (SIC) technology is a key technique. Traditional SIC is typically realized as fixed SIC (FSIC), which inherently suffers from unavoidable error floors in outage probability due to its rigid decoding order \cite{sun2021new}. Therefore, a novel SIC technology called hybrid SIC (HSIC) has been applied to the NOMA transmission phase of the H‑NOMA system \cite{Sun2025HSICnPA}. By dynamically adjusting the decoding order, HSIC outperforms traditional OMA in both transmission rate and energy efficiency. However, this HSIC-aided H-NOMA scheme can avoid outage probability error floors only when user's target rates satisfy strict conditions \cite{sun2023hybrid}. To overcome this limitation, HSIC has been further integrated with power adaptation (PA), enabling the H-NOMA system to achieve superior performance without such restrictive constraints \cite{Wang2025HSICPA, Wang2026CompPerf}. 

Nevertheless, existing studies on H-NOMA adopt an equal power allocation strategy, where the opportunistic user applies the same power reduction coefficient in both its dedicated OMA time slot and the shared NOMA time slot. However, the two time slots exhibit fundamentally different transmission characteristics: in the NOMA slot, the opportunistic user shares the time-frequency resource with the legacy user and is subject to inter-user interference and the HSIC decoding condition, whereas in its dedicated OMA slot, it transmits without such interference. As a result, a uniform power reduction coefficient may not be optimal. This naturally raises a key question: can differentiated power reduction coefficients be configured for the two time slots so as to maximize the achievable rate of the H-NOMA system?

This paper investigates the optimal transmit power allocation technique for the HSIC-PA aided H-NOMA network. Specifically, the main contributions of this work are summarized as follows:
\begin{itemize}
\item We design a differentiated optimal power allocation scheme with distinct power reduction coefficients $\beta_1$ and $\beta_2$ for the OMA and NOMA time slots of H-NOMA transmission mechanism, respectively. Unlike existing studies that adopt a uniform coefficient, the two coefficients are jointly constrained by $\beta_1 + \beta_2 = \eta$, which makes the power allocation across the two slots inherently joint rather than independent. This constraint also ensures that the energy consumption of the proposed scheme is lower than traditional OMA. In addition, HSIC-PA is employed to facilitate signal transmission in the NOMA time slot.

\item We derive the closed-form optimal solution of the power allocation coefficients that maximizes the data rate, along with the corresponding rate expressions, by considering different HSIC decoding modes and feasibility conditions. Simulation results verify the accuracy of the theoretical analysis and show that the proposed scheme achieves better data rate performance than the equal-power allocation scheme, further validating the effectiveness of the proposed scheme.
\end{itemize}

\section{System Model}

We consider an uplink communication system consisting of a base station (BS) and multiple users, where the users are divided into legacy users and opportunistic users. Before H-NOMA transmission, the users are paired according to predefined pairing criteria to form multiple transmission groups, with each group consisting of one legacy user and one opportunistic user. Different transmission groups are allocated orthogonal time-frequency resources, so that inter-group interference is avoided. Existing NOMA studies have also considered such user pairing mechanisms in multi-user systems, where the pairing relationship is typically determined based on user characteristics such as channel conditions \cite{ZhangUserPair2016, ChenUserPair2018}. Given the pairing results, we select one representative transmission group for subsequent analysis, whose users are denoted as \(U_m\) (the legacy user) and \(U_n\) (the opportunistic user). Under the traditional time division multiple access (TDMA) mechanism, each user is allocated a dedicated time slot with duration $T$. We assume that the background noise at each user is normalized to unity, and consider a slowly time-varying channel, which means that the channel gain of each user remains constant within a frame. The Rayleigh fading model is adopted to facilitate closed-form analytical derivations.
Accordingly, the achievable data rate of user $U_i \ (i \in \{m,n\})$ is given by $\log(1+\rho_i|h_i|^2)$, where $\rho_i$ and $h_i$ denote the transmit power and the normalized small-scale Rayleigh fading coefficient of user $U_i$, respectively. For analytical tractability, perfect CSI is assumed at the BS, which is also a common assumption in theoretical NOMA analysis. In practical deployments, however, channel estimation errors, feedback delays, and quantization noise may lead to imperfect CSI, affecting the accuracy of HSIC decoding mode selection and power allocation coefficient calculation. The impact of generalized fading channels and imperfect CSI will be investigated in future work.

In the H-NOMA scheme, $U_m$ can only transmit signals within its allocated time slot (denoted as the ``$m$-th time slot"). In contrast, $U_n$ can not only transmit signals in its own allocated time slot (denoted as the ``$n$-th time slot") but also share the $m$-th time slot for signal transmission without degrading the transmission reliability of $U_m$. 
The preset target rate of $U_m$ is denoted as $R_m$, which is the minimum rate required for successful transmission of $U_m$.
For NOMA transmission in the $m$-th time slot of the H-NOMA scheme, it is essential to ensure that the transmission of $U_n$ does not degrade the transmission reliability of $U_m$ relative to the OMA scheme. 
To address this requirement, the HSIC-PA aided H-NOMA scheme is adopted in this paper. 

Since \(U_n\) adopts different transmission modes in the two slots, it experiences distinct channel conditions. In the NOMA slot, \(U_n\) transmits concurrently with \(U_m\), and its achievable rate is constrained by inter‑user interference and the HSIC decoding conditions. By contrast, \(U_n\) performs interference‑free independent transmission in the OMA slot. Hence, power allocation should be separately considered for these two slots. In the HSIC‑PA aided H‑NOMA scheme, the maximum allocated power for \(U_n\) is \(\beta_1 P_n\) in the NOMA slot and \(\beta_2 P_n\) in the OMA slot, where \(\beta_1\) and \(\beta_2\) denote the power reduction coefficients for the NOMA slot and OMA slot, respectively. The parameters \(\beta_1\) and \(\beta_2\) satisfy $\beta_1+\beta_2=\eta$, where \(0\leq\beta_1\leq1\), \(0\leq\beta_2\leq1\), and \(0<\eta\leq1\). This constraint arises from the total power budget of \(U_n\), which must be shared across both transmission slots, thereby coupling the power allocation between them rather than allowing independent determination. With such a constraint on \(\eta\), the power consumption of the HSIC‑PA aided H‑NOMA scheme is no greater than that of the traditional OMA scheme.

If $U_n$ operates in traditional OMA mode, it transmits data only in the $n$-th time slot, with a data rate of
\begin{equation}
    R_n = \log  \big (1+\rho_n \left| h_n \right|^2 \big ).
\end{equation}

Let $P_i = \rho_i |h_i|^2$ denote the power of $U_i$ received by the BS, where $i = m, n$. Define a threshold $\tau_m$, which represents the maximum interference level at which $U_m$ can still maintain the same outage performance as in the OMA scheme, and is given by:
\begin{align}\label{equ_tau} 
\begin{split}
\tau_m = \max\big(0, \frac{\rho_m |h_m|^2}{2^{R_m} - 1} - 1\big)  
       = \max\big(0, \frac{P_m}{\varepsilon_m} - 1),         
\end{split}
\end{align}
where $\varepsilon_m = 2^{R_m} - 1$. 

For the OMA slot of the H-NOMA scheme, the achievable data rate of $U_n$ is expressed as
\begin{align}
    R^{\text{OMA}} = \log\big(1 + (\eta - \beta_1) P_n\big). 
\end{align}

Based on the relationship between $\beta_1P_n$ and $\tau_m$, the SIC strategies for the NOMA slot of the H-NOMA scheme can be classified into the following two types:
\subsection{Type A: $\beta_1P_n\le\tau_m$}

In this specific scenario, decoding $U_n$ in the second SIC stage not only enhances its individual throughput, but also maintains the interference cancellation order required for $U_m$ to successfully decode its own signal. Therefore, positioning $U_n$ at the second SIC stage is advantageous, and its corresponding achievable rate is expressed as: 
\begin{align}
R_A^{\text{NOMA}} = \log\left(1 + \beta_1 P_n\right)
\end{align}
Hence, the data rate of $U_n$ for the \textit{Type A} scenario is: 
\begin{align}\label{R_A}
\begin{split}
R_A &= R_{A}^{\text{NOMA}} \!+\! R^{\text{OMA}} 
=\log\!\big(1 \!+\! \eta P_n \!+\! P_n^2 \eta \beta_1 \!-\! P_n^2 \beta_1^2 \big)
\end{split}
\end{align}

\subsection{Type B: $\beta_1P_n>\tau_m$}
In this specific scenario, the data rate for $U_n$ is defined in two distinct cases as follows:
\begin{itemize}
\item Case 1: In this case, the power adaptation mechanism is introduced by selecting a power adaptation factor $0< \lambda \leq 1$ such that $\lambda \beta_1 \rho_n \left| h_n \right|^2 = \tau_m$. Consequently, the signal of $U_n$ can be decoded at the second SIC stage, and $U_n$'s data rate in NOMA slot is given by: 
\begin{equation}
    R_{B1}^{\text{NOMA}} =  \log \big( 1+  \tau_m \big).
\end{equation}
\item Case 2: In this case,  the signal of $U_n$ is decoded at the first stage of SIC, and $U_n$'s data rate in the NOMA time slot is given by: 
\begin{equation}
    R_{B2}^{\text{NOMA}} = \log \big( 1 + \frac{\beta_1 P_n}{1+P_m} \big).
\end{equation}
\end{itemize}

In summary, the data rate of $U_n$ for the \textit{Type B} scenario is the maximum of the two cases: 
\begin{align}
    R_{B}^{\text{NOMA}}  
 \!=\!
    \begin{dcases} 
    \!R_{B1}^{\text{NOMA}}, &\tau_m \!>\! \frac{\beta_1 P_n}{1 \!+\! P_m}\\        \!R_{B2}^{\text{NOMA}}, &\tau_m \!<\! \frac{\beta_1 P_n}{1 \!+\! P_m} 
    \end{dcases}
\end{align}

Hence, the achievable data rate of $U_n$ for the two cases of the \textit{Type B} scenario is expressed as $R_{B1}$ and $R_{B2}$, respectively. 
\begin{align}
    R_{B1} &= R_{B1}^{\text{NOMA}} + R^{\text{OMA}} , \quad \tau_m > \frac{\beta_1 P_n}{1 + P_m} 
 \label{R_B1}\\
    &=\log\big((1 + \tau_m)(1 + \eta P_n - \beta_1 P_n)\big)\nonumber \\ 
    R_{B2} &= R_{B2}^{\text{NOMA}} + R^{\text{OMA}} \label{R_B2}, \quad \tau_m < \frac{\beta_1 P_n}{1 + P_m}  \\
    &=\log\Big(1 \!+\! \eta P_n \!+\! \frac{P_n (\eta P_n - P_m)}{1 + P_m} \beta_1 \!-\! \frac{P_n^2}{1 + P_m} \beta_1^2\Big)\nonumber
\end{align}

Moreover, by further analyzing the range of values $\beta_1$ and $\tau_m$ in different scenarios, the data rate of $U_n$ can be summarized as follows:
\begin{align}\label{equ_Rn}
R_n(\beta_1) = 
\begin{dcases} 
R_A, &  \beta_1 \leq \frac{\tau_m}{P_n} \\
R_{B1}, & \frac{\tau_m}{P_n} < \beta_1 < \frac{(1 + P_m) \tau_m}{P_n} \\
R_{B2}, & \beta_1 \geq \frac{(1 + P_m) \tau_m}{P_n} 
\end{dcases}
\end{align}

As shown in the above equation, the data rate is represented as a function of $\beta_1$. To obtain the maximum value of $R_n$, it is necessary to analyze the range of $\beta_1$ and its optimal value.

\section{Optimal Power Allocation Analysis}
Since the inner functions of the logarithmic expressions in \eqref{R_A} and \eqref{R_B2} are downward-opening quadratic functions, and the inner function of \eqref{R_B1} is bounded and monotonically decreasing, a unique maximum value exists within the feasible domain.
By simple calculations, the theoretically optimal value of $\beta_1$ that maximizes the data rate for each scenario is obtained as follows, denoted by $\hat{\beta}_1$:
\begin{align}\label{theor_β1}
\hat{\beta}_{1} = 
\begin{dcases}
\hat{\beta}_{1,A} \ = \frac{\eta}{2}, \\
\hat{\beta}_{1,B1} = \frac{\tau_m}{P_n}, \\
\hat{\beta}_{1,B2} = \frac{\eta}{2} - \frac{P_m}{2P_n}. 
\end{dcases}
\end{align}
According to \eqref{equ_tau}, $\tau_m$ can take two different values. Therefore, further analysis of $\tau_m$ is required  to obtain the actual optimal solutions, denoted by $\beta^*_1$. 
Note that $\hat{\beta}_1$ is the theoretical optimal value obtained without considering the feasible range of $\beta_1$, whereas $\beta^*_1$ is the actual optimal value obtained after accounting for the feasible range imposed by $\eta$ and $\tau_m$.


\subsection{Scenario 1: $\tau_m=0$, i.e., ${P_m}\leq{\varepsilon_m}$}
Since $\beta_1P_n>\tau_m$, $U_n$ can only transmit at the rate of $R_{B2}$ in this case. Based on \eqref{theor_β1} and the condition that $\beta_1 \ge 0$, the actual optimal value of $\beta_1$ is:
\begin{align}
{\beta}^*_{1,B2}=
    \begin{dcases}
     0, &\eta \leq \frac{P_m}{P_n}\\
     \frac{\eta}{2} - \frac{P_m}{2 P_n}, &\eta > \frac{P_m}{P_n}
    \end{dcases}
\end{align}

Therefore, when $\tau_m=0$, the corresponding maximum data rate $R^*_{n}$ is:
\begin{align} \label{equ_RnS1_S2}
\!\!\! R^*_{n} =\! \!
\left\{
\begin{aligned}
&S_1: R_{B2}(0) = \log\left(1+\eta P_n\right), \\[2pt]
&S_2: R_{B2}\Bigl(\frac{\eta}{2}-\frac{P_m}{2P_n}\Bigr) 
      = \log\Bigl(1+\frac{\eta P_n-P_m}{2(1+P_m)}\Bigr) \\
     &\qquad\qquad\qquad\quad\quad\quad \ + \log\Bigl(1+\frac{\eta P_n+P_m}{2}\Bigr).
\end{aligned}
\right.
\end{align}
Here, $S_i$ denotes the $i$-th situation (i.e., the $i$-th globally indexed case across all scenarios) considered in this paper.


\subsection{Scenario 2: $\tau_m>0$, i.e., ${P_m}>{\varepsilon_m}$}

In this scenario, $U_n$ can transmit at data rates $R_A$, $R_{B1}$ and $R_{B2}$ according to different values of $\beta_1$ as shown in \eqref{equ_Rn}. By comparing the value range of $\beta_1$ with the constraint $\beta_1\le1$, we obtain the following three conditions.

\subsubsection{Condition 1: $\frac{\tau_m}{P_n} \ge 1$, i.e., $\tau_m \ge P_n$}  

$U_n$ can only transmit in the $R_A$ mode. The actual value of $\beta_1$ corresponding to the maximum transmission rate of $R_A$ is $\frac{\eta}{2}$. Therefore, the maximum transmission rate in this case is:
\begin{align}
    S_3: R^*_{n} = R_A\left( \frac{\eta}{2} \right) = 2\log\left(1 + \frac{\eta}{2} P_n\right), \\ \beta^*_{1,A} = \frac{\eta}{2}, \ \tau_m \ge P_n. \nonumber
\end{align}
\subsubsection{Condition 2: $\frac{\tau_m}{P_n} < 1 \leq \frac{(1 + P_m)\tau_m}{P_n}$, i.e., $\frac{P_n}{1 + P_m} \leq \tau_m < P_n$}  
The transmission mode of $U_n$ can be $R_A$ or $R_{B1}$.
\begin{itemize}    
\item \textit{Mode} $R_A$: $\beta_1 \leq \frac{\tau_m}{P_n}$
\end{itemize}

In this case, the theoretically optimal value of $\beta_1$ is $\frac{\eta}{2}$. However, constrained by the interval of $\beta_1$, it is necessary to compare the relationship between $\frac{\tau_m}{P_n}$ and $\frac{\eta}{2}$, which is as follows:
\begin{align}
\begin{dcases} 
\tau_m > \frac{P_n \eta}{2},\, \beta_{1,A}^* = \dfrac{\eta}{2}, \\
\tau_m \leq \frac{P_n \eta}{2},\, \beta_{1,A}^* = \dfrac{\tau_m}{P_n}. 
\end{dcases}
\end{align}

Secondly, since $\frac{P_n}{1 + P_m} \leq \tau_m < P_n$ and $\frac{P_n\eta}{2}<P_n$, it is also necessary to compare the relationship between $\frac{P_n}{1 + P_m}$ and $\frac{P_n\eta}{2}$. Based on this comparison, the maximum transmission rate in this case is:
\begin{align}
R^*_{n} =
\begin{dcases}
R_A\left( \frac{\eta}{2} \right), &\beta^*_{1,A} = \frac{\eta}{2},  \; \eta \leq \frac{2}{1 + P_m}, \quad \\  &\frac{P_n}{1 + P_m} \leq \tau_m < P_n,  \\ 
R_A\left( \frac{\eta}{2} \right), &\beta^*_{1,A} = \frac{\eta}{2},  \; \eta > \frac{2}{1 + P_m}   \\ &\frac{P_n \eta}{2}, \leq \tau_m < P_n,\\
R_A\left( \frac{\tau_m}{P_n}\right), &\beta^*_{1,A} = \frac{\tau_m}{P_n}, \; \eta > \frac{2}{1 + P_m}, \\ &\frac{P_n}{1 + P_m} \leq \tau_m < \frac{P_n \eta}{2}.
\end{dcases}
\end{align}

\begin{itemize}
\item \textit{Mode} $R_{B1}$: $\frac{\tau_m}{P_n} < \beta_1 \leq 1$
\end{itemize}

The theoretically optimal solution for $\beta_{1}$ in this case is $\dfrac{\tau_m}{P_n}$. 
According to \eqref{R_B1}, $R_{B1}$ gradually decreases as $\beta_1$ increases within the interval of this case. Thus, the optimal solution for \textit{Mode} $R_{B1}$ is:
\begin{equation}
R^*_n = R_{B1}\left( \frac{\tau_m}{P_n}\right), \quad  \beta^*_{1,B1} = \frac{\tau_m}{P_n}. 
\end{equation}

When considering \textit{Mode} $R_{A}$ and \textit{Mode} $R_{B1}$ together, $R_{B1}(\beta_{1}^*) \le R_A(\beta_{1}^*)$ always holds.
In summary, the maximum rate $R_n^*$ of $U_n$ in Condition 2 is:
\begin{align} \label{equ_RnS4_S5_S6}
R_n^* = 
\left\{
\begin{aligned}
&S_4: R_A\left( \frac{\eta}{2} \right) = 2\log\left( 1 + \frac{\eta P_n}{2} \right),  \\[2pt]
&S_5: R_A\left( \frac{\eta}{2} \right) = 2\log\left( 1 + \frac{\eta P_n}{2} \right),  \\[2pt]
&S_6: R_A\left( \frac{\tau_m}{P_n} \right) = \log\left( 1 + \tau_m \right) \\
     &\qquad\qquad\quad\quad \ \   + \log\left( 1 + \eta P_n - \tau_m \right).
\end{aligned}
\right.
\end{align}

\subsubsection{Condition 3: $\frac{(1 + P_m)\tau_m}{P_n}<1$, i.e., $0 < \tau_m < \frac{P_n}{1 + P_m}$} 
In this condition, the transmission modes can be $R_A$, $R_{B1}$, and $R_{B2}$.
\begin{itemize}
\item \textit{Mode} $R_A$: $\beta_1 \leq \frac{\tau_m}{P_n}$
\end{itemize}

Similar to the analysis in Condition 2, within the range of $0 < \tau_m < \frac{P_n}{1 + P_m}$, the value of $\beta_{1,A}^*$ is:

\begin{equation} \label{A}
\beta^*_{1,A}=
\begin{dcases} 
\frac{\eta}{2}, & \eta \leq \frac{2}{1 + P_m}, \; \frac{P_n \eta}{2} < \tau_m < \frac{P_n}{1 + P_m} \\
\frac{\tau_m}{P_n}, & \eta \leq \frac{2}{1 + P_m}, \; 0 < \tau_m \leq \frac{P_n \eta}{2} \\
\frac{\tau_m}{P_n} ,& \eta > \frac{2}{1 + P_m}, \; 0 < \tau_m < \frac{P_n}{1 + P_m}
\end{dcases}
\end{equation}

\begin{itemize}
\item \textit{Mode} $R_{B1}$: $\frac{\tau_m}{P_n} \leq \beta_1 < \frac{(1 + P_m)\tau_m}{P_n}$
\end{itemize}

Similarly, the actual optimal solution for $\beta_{1}$ in this case is
\begin{equation} \label{B1}
\beta_{1,B1}^* = \dfrac{\tau_m}{P_n}   
\end{equation}

\begin{itemize}
\item \textit{Mode} $R_{B2}$: $\frac{(1 + P_m)\tau_m}{P_n} \leq \beta_1 \leq 1$
\end{itemize}

In this case, the theoretically optimal value of $\beta_1$ is $\frac{\eta}{2}-\frac{P_m}{2P_n}$. It is necessary to compare the relationship between $\frac{(1 + P_m)\tau_m}{P_n}$ and $\frac{\eta}{2}-\frac{P_m}{2P_n}$, which is as follows:
\begin{align}
\begin{dcases} 
\tau_m \geq \frac{P_n \eta - P_m}{2(1 + P_m)},\; \beta_{1,B2}^* = \dfrac{(1 + P_m)\tau_m}{P_n} \\
 \tau_m < \frac{P_n \eta - P_m}{2(1 + P_m)},\; \beta_{1,B2}^* = \frac{\eta}{2} - \frac{P_m}{2P_n} 
\end{dcases}
\end{align}


Secondly, given that $0 < \tau_m < \frac{P_n}{1 + P_m}$, it is necessary to compare the three quantities: $\frac{P_n}{1 + P_m}$, $\frac{P_n \eta - P_m}{2(1 + P_m)}$, and $0$.

We first compare $\frac{P_n \eta - P_m}{2(1 + P_m)}$ with $\frac{P_n}{1 + P_m}$. It is straightforward to show that $\frac{P_n}{1 + P_m} > \frac{P_n \eta - P_m}{2(1 + P_m)}$. Then, we compare $\frac{P_n \eta - P_m}{2(1 + P_m)}$ with $0$. Based on this comparison, the value of $\beta_{1,B2}^*$ can be expressed as follows:
\begin{align}\label{B2}
\beta_{1,B2}^* \! = \!
\begin{dcases} 
\frac{(1 + P_m)\tau_m}{P_n},& \eta > \frac{P_m}{P_n}, \\ &\frac{P_n \eta \!-\! P_m}{2(1 \!+\! P_m)} \!\leq\! \tau_m \!<\! \frac{P_n}{1 \!+\! P_m}, \\
\frac{\eta}{2} - \frac{P_m}{2P_n}, & \eta > \frac{P_m}{P_n}, \\ &0 < \tau_m < \frac{P_n \eta - P_m}{2(1 + P_m)}, \\
\frac{(1 + P_m)\tau_m}{P_n},& \eta \leq \frac{P_m}{P_n}, \\ &0 < \tau_m < \frac{P_n}{1 + P_m},
\end{dcases}
\end{align}

Through a comprehensive analysis of \eqref{A}, \eqref{B1} and \eqref{B2}, the maximum data rate $R_{n}^*$ of $U_n$ in Condition 3 can be expressed as \eqref{equ_RnS7_S8_S9}. 
\begin{align} \label{equ_RnS7_S8_S9}
R_n^* = 
\begin{dcases} 
S_7: R_A\left( \frac{\eta}{2} \right), & \ 0 < \eta \leq \frac{2}{1 + P_m},\\&\ \frac{P_n \eta}{2} < \tau_m < \frac{P_n}{1 + P_m}, \\
S_8: R_{B2}\left( \frac{\eta}{2} - \frac{P_m}{2P_n} \right), &\ \frac{P_m}{P_n} < \eta \leq 1,\\&\ 0 < \tau_m \leq \tau^-_{m0}, \\
S_9: R_A\left( \frac{\tau_m}{P_n} \right),\ & \ \text{others}
\end{dcases}    
\end{align} 

In summary, the conclusions for all cases are summarized in Table \ref{tab_OptimalValue}. Although the derivation of the optimal power allocation coefficient is relatively involved, the resulting closed-form solution only involves a small number of basic arithmetic operations. Based on the instantaneous channel conditions, the corresponding optimal power allocation coefficient can be directly obtained in closed form. Therefore, the proposed scheme incurs negligible computational complexity and limited additional signaling overhead in practical implementation.

\begin{table*}[!ht]
    \centering\renewcommand\arraystretch{2}
    \caption{The optimal values of $\beta^*_1$ and the maximum data rates $R^*_n$ for different situations.}
    \label{tab_OptimalValue}
    \begin{tabular}{|c|c|c|c|c|c|c|}
\hline \multicolumn{2}{|c|}{Scenario} & Situation & $R_n^*$ & $\beta_1^*$ & $\tau_m$ & $\eta$ \\
\hline 
\multicolumn{2}{|c|}{\multirow{2}{*}{
\makecell{Scenario 1:
$\tau_m=0$ \\
 (i.e., $P_m \leq \varepsilon_m $ ) }
}} & $S1$ & $R_{B2}(0)$&0 &$\tau_m=0$ & $\eta \leq \frac{P_m}{P_n}$\\\cline{3-7}
\multicolumn{2}{|c|}{}  & $S2$ & $R_{B2}\left( \frac{\eta}{2} \!-\! \frac{P_m}{2 P_n} \right)$& $\frac{\eta}{2} -\frac{P_m}{2 P_n}$& $\tau_m=0$&$\eta > \frac{P_m}{P_n}$ \\\hline
\multirow{7}{*}{\multirow{1}{*}{
\makecell{Scenario 2:
$\tau_m>0$ \\
 (i.e., $P_m > \varepsilon_m $ )}}}&\renewcommand\arraystretch{1}
 \multirow{1}{*}{\makecell{Condition 1: \\ $\tau_m \ge P_n$}}
& $S3$ & $R_A\left( \frac{\eta}{2} \right)$& $\frac{\eta}{2}$& $\tau_m \ge P_n$& $0<\eta\leq 1$ \\\cline{2-7}
 && $S4$ &$R_A\left( \frac{\eta}{2} \right)$ &$\frac{\eta}{2}$ & $\frac{P_n}{1 + P_m} \leq \tau_m < P_n$&$\eta \leq \frac{2}{1 + P_m}$\\\cline{3-7}
 &\renewcommand\arraystretch{1}
 \multirow{1}{*}{
 \makecell{Condition 2: \\ $\frac{P_n}{1 + P_m} \leq \tau_m < P_n$}}
& $S5$ & $R_A\left( \frac{\eta}{2} \right)$& $\frac{\eta}{2}$& $\frac{P_n \eta}{2} < \tau_m < P_n$& $\eta > \frac{2}{1 + P_m}$\\\cline{3-7}
 && $S6$ & $R_A\left( \frac{\tau_m}{P_n} \right)$& $ \frac{\tau_m}{P_n}$&$\frac{P_n}{1 + P_m} \leq \tau_m \leq \frac{P_n \eta}{2}$ & $\eta > \frac{2}{1 + P_m}$\\\cline{2-7}
 && $S7$ & $R_A\left( \frac{\eta}{2} \right)$& $\frac{\eta}{2}$&$\frac{P_n \eta}{2} < \tau_m < \frac{P_n}{1 + P_m}$ & $0 < \eta \leq \frac{2}{1 + P_m}$\\\cline{3-7}
 &\renewcommand\arraystretch{1}
  \multirow{1}{*}{
\makecell{Condition 3: \\ $0 < \tau_m < \frac{P_n}{1 + P_m}$} }
& $S8$ &$R_{B2}\left( \frac{\eta}{2} - \frac{P_m}{2P_n} \right)$&$\frac{\eta}{2} - \frac{P_m}{2P_n}$& $\ 0 < \tau_m \leq \tau^-_{m0}$&$\frac{P_m}{P_n} < \eta \leq 1$ \\\cline{3-7}
 && $S9$&$R_A\left( \frac{\tau_m}{P_n} \right)$ & $\frac{\tau_m}{P_n}$& others& others \\\hline 
\end{tabular}
\end{table*}

\vspace{-0.5em}
\section{Numerical Results}
In this section, numerical results are provided to validate the derived analytical results and demonstrate the superior performance of the proposed scheme. The channel gains $h_m$ and $h_n$ are identically distributed Rayleigh fading with the same variance. The maximum transmission power of $U_m$ and $U_n$ is equal, i.e., $\rho_m = \rho_n$. With the noise power normalized to unity, the SNR in this section is simply $\rho_n$.

\begin{figure}[t!]
	\centering
\subfloat
{\includegraphics[width=1\linewidth]{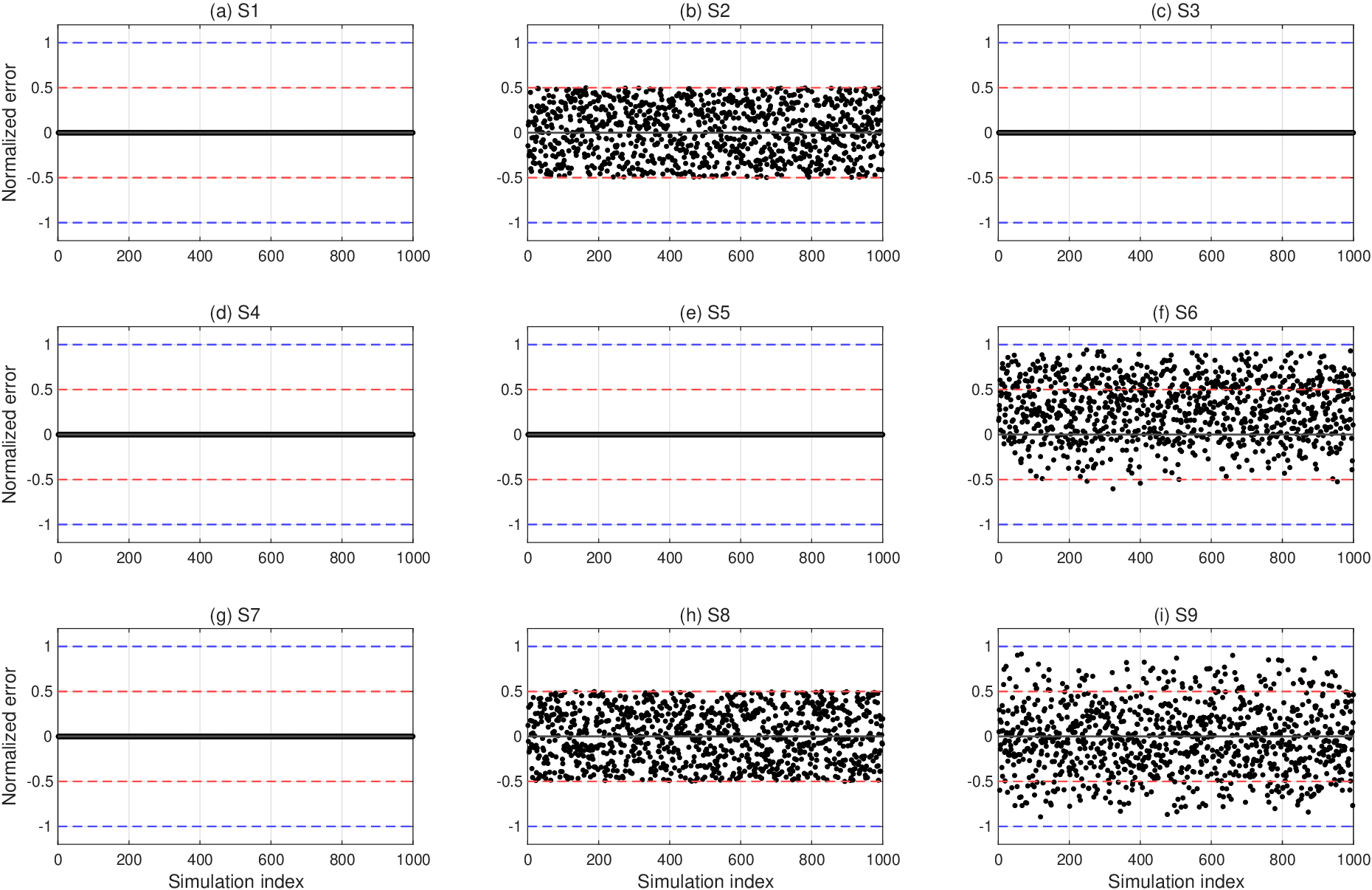}}
\caption{
Normalized error for cases $S1$–$S9$ over 1000 independent simulation trials. The solid black line denotes zero error. The red and blue dashed lines denote the $\pm \Delta/2$ and $\pm \Delta$ bounds, respectively. $\rho_n = \rho_m$, $R_m = 2$, $\eta=0.2$.}
\label{fig1}
\vspace{-1em}
\end{figure}

Fig.~\ref{fig1} presents the normalized error $e/\Delta = (\beta^*_{\text{sim}}-\beta^*_{\text{ana}})/\Delta$ for the nine cases over 1000 independent simulation trials, where $\beta^*$ denotes the optimal power allocation factor, $\Delta$ is the search step size for $\beta$. Here,  $\beta^*_{\text{sim}}$ and $\beta^*_\text{ana}$ are the simulated and analytical values of $\beta^*$, respectively. The results can be categorized into three groups according to the characteristics of the  solution. For $S1$, $S3$, $S4$, $S5$ and $S7$, $\beta^*_\text{ana}$ is a constant that can be exactly reached by the discrete search step, resulting in zero error. For $S2$ and $S8$, $\beta^*_\text{ana}$ corresponds to the extrema of a continuous smooth function, and the error is bounded within $\pm \Delta/2$. For $S6$ and $S9$, the optimum lies near the transition point of a piecewise non-smooth function where the derivative is discontinuous, and the resulting error is bounded within $\pm \Delta$. These results not only validate the analytical derivations but also reveal how the properties of the objective function affect the search accuracy. Overall, the simulation results closely match the analytical results, confirming the accuracy of the derived closed-form solution.

\begin{figure}[t!] 
    \centering
    \vspace{-1em}
    \setlength{\abovecaptionskip}{0em}   
    \setlength{\belowcaptionskip}{0em}   
    \subfloat[${\rho_n}={\rho_m}$, $\eta=0.8$]
    {\includegraphics[width=0.8\linewidth]{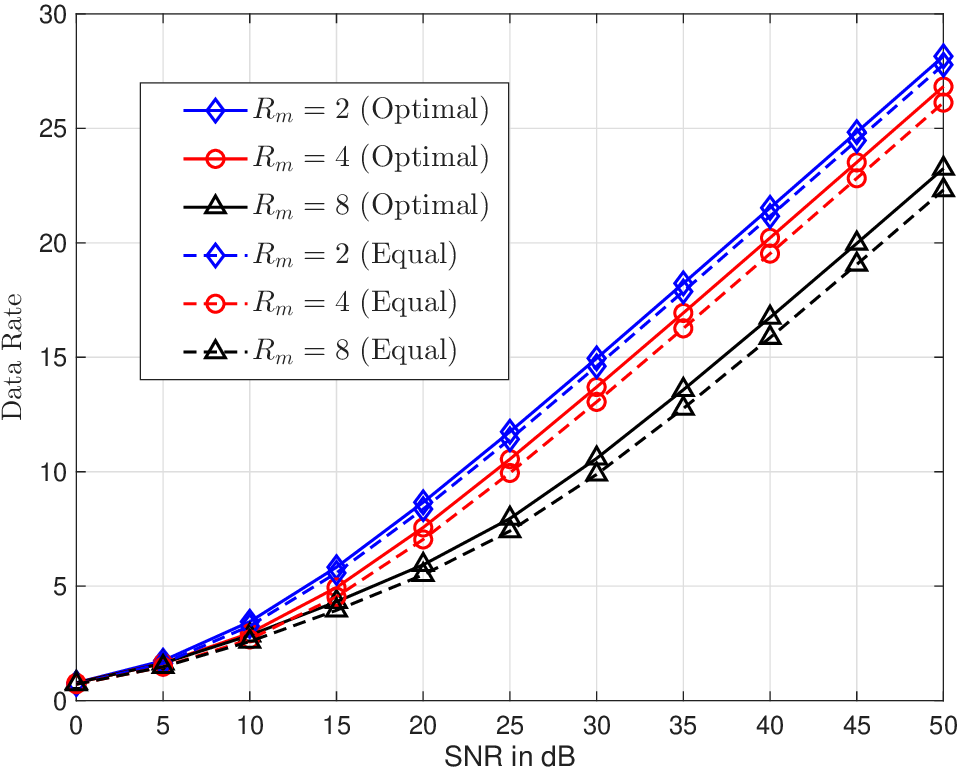}}\\
    \subfloat[${\rho_n}={\rho_m}$, $R_m=8$]
    {\includegraphics[width=0.8\linewidth]{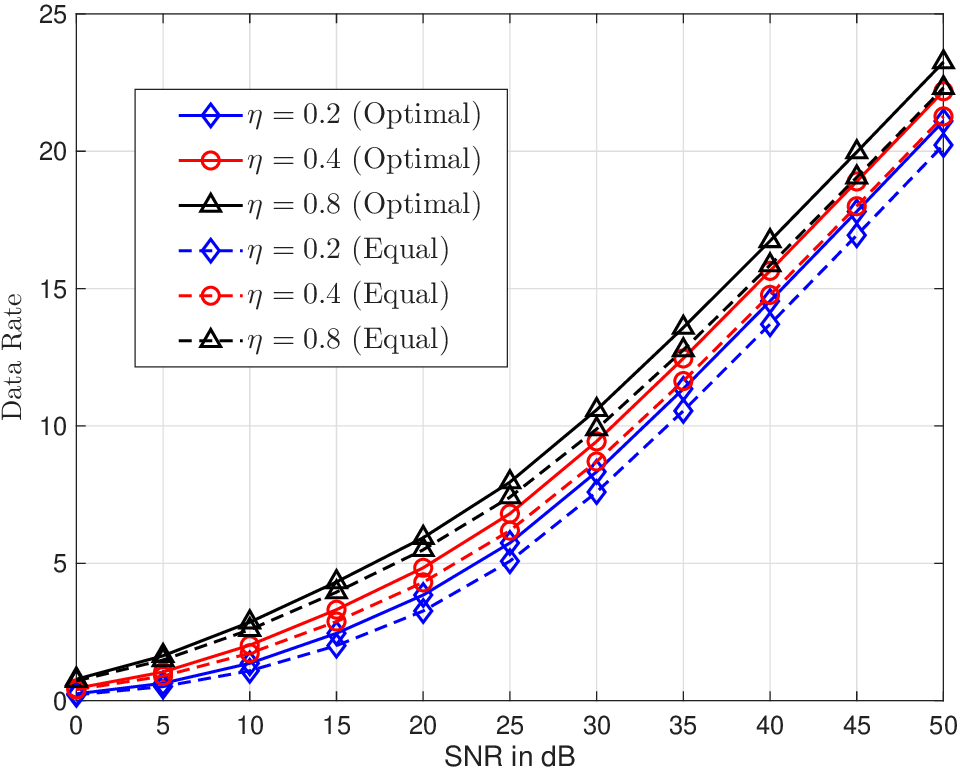}}
    \caption{Comparison of data rate between optimal power allocation and equal power allocation}
    \label{fig2}
    \vspace{-1em}
\end{figure}

Fig. \ref{fig2} shows the performance characteristics of the proposed H-NOMA scheme. Note that the total resource allocation coefficient $\eta$ is less than 1, indicating that the system power consumption is lower than that of traditional OMA. The results show that the optimal power allocation scheme achieves better performance than the equal power allocation strategy. This performance gain further demonstrates the benefit of cross-slot joint power allocation, where the power reduction coefficients $\beta_1$ and $\beta_2$ are jointly optimized under the constraint $\beta_1 + \beta_2 = \eta$, rather than being determined independently.

\begin{figure}[ht!]
	\centering
	\vspace{-1em}
	\setlength{\abovecaptionskip}{0em}  	
        \setlength{\belowcaptionskip}{0em}   
\subfloat
{\includegraphics[width=0.8\linewidth]{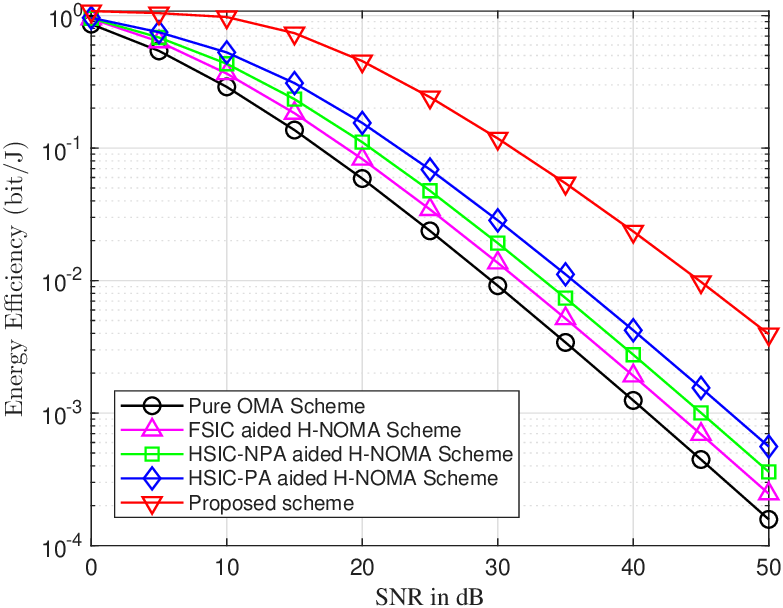}} 
\caption{\color{black}Energy efficiency comparison of different schemes. $\rho_n = \rho_m$, $R_m = 2$, $\eta=0.6$}
\label{fig3}
\vspace{-1em}
\end{figure}

To further evaluate the performance of the proposed scheme, simulations are conducted from the perspective of energy efficiency, which is defined as the achievable data rate per unit actual transmit power. For the opportunistic user \(U_n\), the energy efficiency is calculated as \(E_p={R_n}/{P_{n,\mathrm{act}}},\) where \(R_n\) denotes the achievable data rate of \(U_n\), and \(P_{n,\mathrm{act}}\) denotes its actual transmit power determined according to the corresponding transmission case and power allocation strategy. Specifically, for the cases without additional power adaptation, the actual total transmit power is $P_{n,\mathrm{act}}=(\beta_1+\beta_2)\rho_n=\eta\rho_n$, whereas for the Type B1 case with the power adaptation factor \(\lambda\), it is $P_{n,\mathrm{act}}=(\lambda\beta_1+\beta_2)\rho_n.$
Fig. \ref{fig3} compares the energy efficiency of five schemes: the pure OMA scheme, the FSIC-PA aided H-NOMA scheme, the HSIC-NPA aided H-NOMA scheme, the HSIC-PA aided H-NOMA scheme, and the proposed scheme. The results show that the proposed scheme achieves the highest energy efficiency over the entire SNR range, further demonstrating the performance advantage of the proposed strategy.

\vspace{-0.5em}
\section{Conclusion}

In this paper, the performance of the HSIC-PA aided H-NOMA scheme has been rigorously evaluated by considering the different transmission conditions in the NOMA and OMA slots. A new analytical characterization of H-NOMA under the considered energy constraint has been developed for the varying-channel scenario, and closed-form optimal power allocation solutions have been derived. The proposed scheme is particularly applicable to cellular and IoT scenarios with heterogeneous QoS requirements, such as industrial monitoring and inspection applications, where a legacy user requires reliable delivery of small data packets and an opportunistic user seeks to maximize its data throughput. While perfect CSI and Rayleigh fading are assumed for analytical tractability, extending the present framework to imperfect CSI and generalized fading channels is a promising direction for future work. In addition, comprehensive comparisons with other power allocation strategies, such as water-filling and energy efficiency-oriented algorithms, will be investigated in future work.

\bibliographystyle{IEEEtran}
\bibliography{IEEEabrv, ref20260401}
\end{document}